\documentclass[a4paper,11pt]{article}
\usepackage{jcappub} % for details on the use of the package, please see the JINST-author-manual
\usepackage{lineno}
\usepackage[utf8]{inputenc}
\usepackage{mathabx}
\usepackage{graphicx}
\arxivnumber{2405.19472}

\title{Implications of in-ice volume scattering for radio-frequency neutrino experiments}
\author{A. Nozdrina$^{1}$ and D. Besson$^{1}$}\affiliation{$^{1}$University of Kansas, Dept.\ of Physics and Astronomy, Lawrence, KS 66045, USA}

\emailAdd{alisa\_nozdrina@ku\_edu}
\abstract{Over the last three decades, several experimental initiatives have been launched with the goal of observing radio-frequency signals produced by ultra-high energy neutrinos (UHEN) interacting in solid media. Observed neutrino event signatures comprise impulsive signals with duration of order the inverse of the antenna+system bandwidth ($\sim$10 ns), superimposed upon an incoherent (typically white noise) thermal noise spectrum. Whereas bulk volume scattering (VS) of radio-frequency (RF) signals is well-studied within the radio-glaciological communities, polar ice-based neutrino-detection experiments have thus far neglected VS in their signal projections. As discussed herein, coherent volume scattering (CVS, for which the phase of the incident signal is preserved during scattering) generated by in-ice neutrino interactions may similarly produce short-duration signal-like power, albeit with a slightly extended time structure, and thereby enhance neutrino detection rates, whereas incoherent (randomized phase) volume scattering (IVS) will persist for O(100 ns), appearing similar to thermal white noise and therefore reducing the measured Signal-to-Noise Ratio (SNR) of neutrino signals. Herein, we present the expected voltage profiles resulting from in-ice volume scattering as a function of the molecular scattering cross-section, for both CVS and IVS, and assess their impact on UHEN experiments. VS contributions are currently only weakly constrained by extant data; stronger limits may be obtained with dedicated calibration experiments.}
\begin{document}
\maketitle

% \begin{abstract}
% %The design for the \emph{Journal of Glaciology} has been implemented as a \LaTeXe\ class file and is derived from article.cls. We recommend that authors use this guide as a template. While writing we suggest you use the two-column \texttt{[twocolumn]} option to check that mathematical equations fit the measure. Submitted papers must, however, be presented using the one-column \texttt{[review]} option. If you have any problems using the class file, plenatbibase contact Overleaf support at \texttt{www.overleaf.com/contact}. The abstract should be less than 200 words and one paragraph long.

%{\huge\textcolor{red}{TBD: Add E conservation check; extension of r to large distances using fast MC, comment on VS in SLAC tests - different VS mechanisms, and limited volume}\textcolor{purple}{https://ieeexplore.ieee.org/document/8453611 - Mie scattering regime / also https://agupubs.onlinelibrary.wiley.com/doi/full/10.1029/2023EA003013}}

\section{~Introduction}
Within the last decade, high energy neutrino astronomy has evolved from high-statistics measurements of atmospheric neutrinos to more recent measurements of the extraterrestrial neutrino flux at TeV$\to$PeV energies\cite{abbasi2021icecube,icecube2022detection,icecube2023observation,icecube2022evidence,icecube2018multimessenger,icecube2021detection}. Ultra high energy neutrinos (UHEN), beyond the Glashow resonance at $E_\nu\sim$6.2 PeV, have too low a flux to be detected by optical techniques. Such neutrinos can arise from decay of charged pions photo-produced by collisions of ultra-high energy cosmic rays with the cosmic microwave background (CMB). Photoabsorption of CMB photons by UHECR is particularly enhanced at center-of-mass energies corresponding to the $\Delta(1232)$ resonance, resulting in both muon and electron neutrinos via the chain $p+\gamma_{CMB}\to\Delta\to n\pi^+;~\pi^+\to\mu{\overline\nu_\mu};~\mu\to e{\overline\nu_e}\nu_\mu$.
They are therefore an unavoidable consequence of both hadronic interactions near the production point, as well as electromagnetic interactions during propagation of the ultra-high energy cosmic rays (UHECR) that have been observed by Auger and the Telescope Array\cite{deligny2023science,kuznetsov2023uhecr}.  
Such neutrinos are also expected as the complement to high-energy cosmic gamma rays produced by neutral pion decay\cite{Cao:2023mig}. 

Many UHEN experiments\cite{Kravchenko:2011im,GorhamAllisonBarwick2009,Barwick:2014rca,aguilar2021design,prohira2021radar,ANITA:2008mzi} have coalesced around the prospect of measuring signals produced by neutrinos interacting in cold polar ice, primarily owing to the excellent transparency of ice to propagating electromagnetic radiation, particularly at radio-frequency wavelengths (${\cal O}$(1~km))\cite{barwick_besson_gorham_saltzberg_2005,aguilar2022radiofrequency}. Two polar detection strategies have shown considerable promise. For the first, Askaryan radiation (generated either in-ice or as the result of the hadronic decay of a $\tau$-lepton in-air) produced by the electron/positron charge imbalance in a shower following a charged- or neutral-current neutrino-ice interaction\cite{Kravchenko:2011im,GorhamAllisonBarwick2009,Barwick:2014rca,aguilar2021design,ANITA:2008mzi,TAROGE:2022soh} produces a measurable RF impulse. For the second, one measures the radar echo produced by the (approximately stationary) ionization trail associated with the shower\cite{prohira2021radar}. 

Projected experimental detection rates are estimated by comparing the calculated signal amplitudes and shapes (aka `templates') against the irreducible noise background (typically, black-body thermal noise from the environment) to which a typical experiment is subject. Since the neutrino flux is a rapidly falling power-law function of energy\cite{ahlers2012minimal}, suppressing the thermal noise background (e.g., by minimizing the noise figure of front-end low-noise amplifiers), or, alternatively, reducing the event trigger threshold relative to thermal noise (e.g., by machine learning or phased array triggering techniques\cite{anker2022improving,Allison:2018ynt}) can have dramatic impact on estimated event detection rates. A reduction in the effective signal-to-noise ratio threshold by a factor of two, for example, corresponds to nearly an order-of-magnitude increase in the detected event rate at the lowest detectable neutrino energies, given the cubic fall-off of the neutrino energy spectrum. 
  
Detection of muon neutrinos in optical experiments is based on the measured arrival time of the first Cherenkov photon incident on a photomultiplier tube (PMT). For the optical detection experiments, both absorption, as well as scattering of Cherenkov photons between production point and the front-end PMT dictate the neutrino detection rates. %The timing and number of photons measured in response to a dedicated light source (LED), e.g., can be used to quantify the two effects. 
At South Pole\cite{aartsen2013measurement}, for example, in-ice scattering lengths vary from 15--60 m, depending on local dust concentrations. The longer absorption lengths (60--200 m) therefore play less of a role in limiting $\nu_\mu$ neutrino detection rates. Water-based optical experiments show an opposite pattern, with the loss of photons dominated by scattering rather than absorption\cite{ANTARES:2004kfl}. 
%THE TIME DEPENDENCE OF THE OPTICAL SIGNAL FOR EXPERIMENT XX IS SHOWN IN FIGURE XXXX
%FIGURE 6 FROM https://arxiv.org/pdf/astro-ph/9903342.pdf
%FIGURE 3 FROM https://www.epj-conferences.org/articles/epjconf/pdf/2016/11/epjconf-VLVnT2015_06009.pdf
%FIGURE 4 FROM https://agupubs.onlinelibrary.wiley.com/doi/full/10.1029/2005JD006687

To quantify scattering in the optical experiments, 
LED flashers are used to determine the time structure of photons which scatter and then are measured by PMTs. We expect the detected photon number to grow with time as the collection volume grows as the cube of distance; eventually, the detected photon number begins to decrease when the photon path length approaches one optical absorption length. For a separation distance {\tt d} between LED and PMT of 75 meters, the total number of scattered photons in IceCube was observed to reach a maximum approximately 75 ns after the arrival of the first photon\cite{ackermann2006optical}; doubling {\tt d} approximately doubles the time for the scattered photon distribution to reach maximum. The (approximately Rayleigh) distribution has a long tail, consistent with multiple scattering effects. Similar distributions were measured at Mediterranean water KM3NET\cite{maragos2016measurement} sites.
%Determinations of radio-frequency transparency do not (thus far) distinguish between absorption vs. bulk ice (volume) scattering and only measure the combined effect in quantifying signal loss.

The Askaryan signal sought for by extant radio experiments has a sharp (O(1ns)), impulsive leading edge, characteristic of the traversal of the Cherenkov shock front across the front-end receiver (Rx) antenna.
In contrast to single photons,
measured broad-band radio signals are temporally extended by the frequency-dependent group delay and the limited antenna and system frequency bandwidth. Consequently, any scattered radio photons that may be present will superpose upon the longer-duration band-limited observed radio signals. In principle, such scattering effects might have been observable in the experimental laboratory tests of the Askaryan effect\cite{ANITA:2006nif,Saltzberg:2000fk}, although the 2-3 meter scale of the laboratory targets limited the radio photon collection volume and the target differed in composition and purity from polar ice. 
To date, {\it in~situ} measurements of RF ice attenuation lengths have not distinguished between true signal absorption and scattering -- the signal amplitude at a receiver point is compared to a transmitted amplitude, and amplitude losses beyond 1/r typically interpreted as due to absorption.

Considerable volume scattering work within the radioglaciological community notwith\-standing\cite{mon1982backward,kim2007scattering,bohren1982radar,wang2016radar,jol2008ground}, neutrino signal estimates have generally not addressed contributions possibly resulting from bulk volume scattering. In the case of the Askaryan experiments, IVS of primary radiation emitted by the shower could potentially bathe the signal in an additional background after the initial neutrino-induced direct signal onset. Since Askaryan radiation is collimated into a Cherenkov cone with approximately 1 degree transverse width, antennas off the Cherenkov-angle may also be subject to an overall enhancement in the long-wavelength photon yield, resulting from the isotropic 
re-scattering from molecules illuminated at, or near the Cherenkov angle. If VS is large enough, this could result in events with an {\cal O}(0.01--1 $\mu$s) duration increase in voltages for times subsequent to the first arrival of the Askaryan pulse, against which the Askaryan signal must be distinguished.

In contrast to the Askaryan detection experiments,
the radar neutrino 
experiments use a clever carrier-cancellation technique to minimize interference from self-induced backgrounds at the carrier frequency\cite{RadarEchoTelescope:2021rca}. This is done by empirically determining the amplitude of a sinusoid injected into the receiver signal path that destructively interferes with the known transmitted carrier continuous wave CW, allowing dynamic suppression of any CW contamination. In this case, only incoherent volume scattering might contribute since any CVS should be suppressed by the (constantly active) carrier cancellation. This technique, however, would not eliminate any CVS produced by reflection of the sounding radar from the sought-after neutrino-induced cascade.

%https://physics.byu.edu/faculty/colton/docs/phy442-winter20/lecture-11-lorentz-oscillator-model.pdf
%https://phys.libretexts.org/Bookshelves/Electricity_and_Magnetism/Essential_Graduate_Physics_-_Classical_Electrodynamics_(Likharev)/07%3A_Electromagnetic_Wave_Propagation/7.02%3A_Attenuation_and_Dispersion
\section{~Additional Experimental Considerations}
Current experimental evidence points out apparent deficiencies in our understanding of radio-wave propagation through ice, which help motivate our current VS study, including:
\begin{itemize}
\item
Previous publications provided experimental evidence for both fixed-frequency (CW), and also broadband signals propagating between two points that, for a smoothly varying refractive index profile dependence on depth, should be `shadowed' relative to each other\cite{Barwick:2018rsp,Deaconu:2018bkf}. This effect was observed, consistently, in data accumulated by the RICE (using CW), ARIANNA (using impulsive signals) and RNO-G (using impulsive signals) experimental groups to fully parameterize the {\it in situ} radio-frequency ice response. The nominally expected shadowing can be considered as a total cancellation of the Huygens wavelets arriving at a given time, propagating from transmitter to receiver. For RICE, waveform captures were observed to be sinusoids, at the carrier frequency, but with significantly reduced amplitude compared to expectations for non-shadowed propagation. The weak shadow zone signals observed in the RICE data were accentuated after co-adding waveforms, indicating that the phase was coherently preserved trigger-to-trigger. By contrast, incoherent volume scattering disrupts a sinusoidal signal at the carrier, due to the `scrambling' of phases %for $\lambda>>d_{scatterer}$ 
and would be expected to simply result in an enhancement of the ambient noise level. %, outside the time window of the signal, as well as some loss of phase coherence in the time window corresponding to observation of the carrier. 

\item Published measurements\cite{Allison:2019rgg} made using a transmitter (Tx) pulser broadcasting within an ice borehole to the englacial ARA radio receiver array show unexpectedly large measured HPol (s-polarization) power relative to VPol (p-polarization) for a predominantly VPol transmitter. Corrected for antenna gain, in some cases the observed HPol power exceeded that measured in VPol; some of this may be attributed to ice birefringence\cite{Heyer:2022ttn,connolly2021impact}. %and ii) the inconsistency between the implied refractive index profiles (with depth) favored by antennas sampling ice at depths of 50 meters vs. antennas sampling ice at depths of 150 meters.
\end{itemize}
These experimental observations indicate that our current understanding of the radio-frequency properties of glacial ice is incomplete, and, in part, prompts our search for evidence of volume scattering.

\section{~Previous calculations and modeling}
\subsection{~The Drude-Lorentz oscillator model}
Electromagnetic scattering can broadly be considered as the atomic response, and re-radiation of an incident electric field.
Modeling an atomic electron as an oscillator bound to a nucleus, the complex permittivity $\epsilon_r$ describing its response to a driving excitation with frequency $\omega_o$ follows directly from ${\vec F}={\rm m}{\vec a}$, leading to:
$$\epsilon_r=1+\omega_p^2/(\omega_0^2-\omega^2-i\omega\gamma),$$
with $\omega_p=\sqrt{Nq^2/(m\epsilon_0)}$ the resonant plasma frequency and $\gamma$ describing collisional damping. The (absorptive) imaginary term in this expression introduces a $\pi/2$ phase shift between the driving term and the (real) scattering response. For a pure medium with no thermal fluctuations, coherent non-dissipative scattering is manifest as the refractive index. In what follows, `volume scattering' refers to the more realistic case of impurities, and/or density fluctuations and/or grain boundaries, etc. If, rather than $\pi/2$, the scattered phase assumes some random value, the incoherent volume scattering (IVS) case obtains; this might arise, e.g., if, under the vertical gravitational stress and the lateral ice-flow-induced strain (which leads to birefringent effects) individual ice grains acquire phase shifts which vary from site to site. In what follows, we retain two primary features (polarization and re-radiation beam pattern) of the Drude-Lorentz model, which also forms the basis for the estimated single-electron Thomson free-particle scattering cross-section $\sigma_T\sim 6.65\times 10^{-25}~\text{cm}^2$.
%mω^2x=9e9*1.6e-19*1.6e-19/(5.29e-11)^2, so ω=sqrt(9e9*1.6e-19*1.6e-19/(9.1e-31*(5.29e-11)**3))=4.13e16; Google says the resonant frequency of ice is around 2 GHz (microwaves); in any case, both optical (blue sky) and radio (ice) are below relevant resonant frequencies

%VS vs. n: https://www.oceanopticsbook.info/view/scattering/physics-scattering

%ZZZ: 1e-5=5.1×10−31*2×1025 - here we want 1e-3 scattered per meter, number of molecules in 1 cubic meter=0.917*6.02e23*1 mole/0.018 ml * 1e6 ml/liter=3.06e28 molecules/cubic meter, so 1e-3=σN and σ=3.27e-32 m^2 or 3.27e-36 cm^2.
\subsection{~Approximate Scale of VS Cross-section}
Perhaps the most familiar example of volume scattering is Rayleigh scattering of visible light in the atmosphere, which corresponds to an in-air scattered power fraction of roughly 1/100,000 per meter travel at optical frequencies. The difference between solid ice and gaseous nitrogen notwithstanding, we consider how the atmospheric (at STP)  Rayleigh cross-section of nitrogen ($5.1\times 10^{-27}~\text{cm}^2$, at optical frequencies) might scale to radio. Unlike our case of dense solid ice (including mobile H+ impurities), the atmospheric cross-section is derived for in-air scattering between sparsely located scattering centers: $\sigma=(2\pi^5d^6/3\lambda^4)((n^2-1)/(n^2+2))^2$ [${\rm m}^2]$, with d the scatterer size (we take 0.3 nm as the size of a water molecule and n=1.78). 
Extrapolated to long-wavelengths, the quartic dependence on $d/\lambda$ therefore suppresses this cross-section by many orders of magnitude (and in fact, well below the cross-section limit we later derive below). Qualitatively, in the Rayleigh model, for which incident signal is re-radiated by atomic dipole `antennas', the small value obtained in this case reflects the scale mismatch between the macroscopic incident signal and the microscopic atomic dipole antenna, as noted in other estimates of VS\cite{romero2024feasibility}. If, however, scattering occurs at the grain, rather than the molecular level, the relevant scattering cross-section is increased by many orders of magnitude.

We can make a dimensional estimate for the maximum scattering cross-section by setting the inverse of the measured radio-frequency attenuation length in cold polar ice ($\sim$800 m)\cite{aguilar2022situ,aguilar2022radiofrequency} equal to the product of the per-ice-molecule VS scattering cross-section multiplied by the molecular number density {\bf n}, given by $917~{\rm kg/m^3} \times 1~{\rm mole/18~ml} \times {\rm 1000 ml/liter} \times 6.02\times 10^{23}$=$3.07\times 10^{28}/{\rm m^3}$. Equating 800 m to 1/$\sigma_{VS}${\bf n} therefore gives $\sigma_{VS}\leq 3.7\times 10^{-28}~\text{cm}^2$. We interpret this value as an upper limit which our more sophisticated calculation should observe.

\subsection{~Previous calculations}
\begin{figure}
\includegraphics[width=\textwidth]{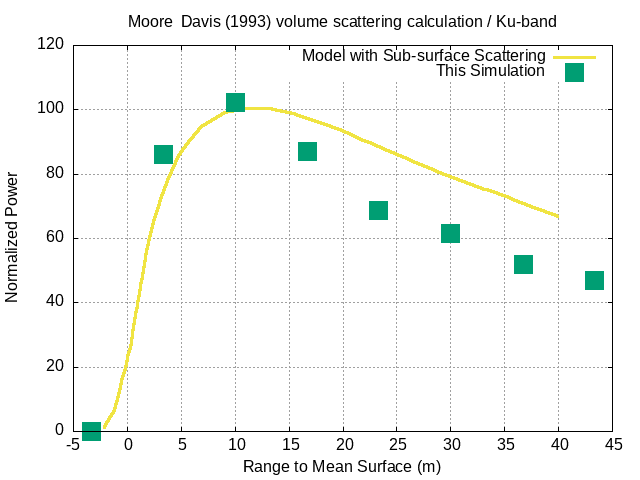}
\caption{Moore \& Davis calculation (yellow), including volume scattering, compared with our simulations (detailed below, and shown as green points).}
\label{fig:MD}
\end{figure}

Davis and Moore\cite{davis_moore_1993} 
%https://www.cambridge.org/core/journals/journal-of-glaciology/article/combined-surfaceand-volumescattering-model-for-icesheet-radar-altimetry/ABC84525CEAB1243BF906A1A01A591B8#R40
previously considered the relative contributions of volume to surface scattering, in the context of satellite- or airplane-based radar surveys of the polar ice sheets, at frequencies in the C- and Ku-bands, and spanning the range 4--17 GHz. Their work followed the observation by Ridley and Partington\cite{partington1989observations,ridley1988model} that an exclusively-surface scattering model was inadequate to explain the total reflected power observed in satellite-based radar measurements. They performed an analytic calculation of the volume scattering fraction of received power measured in the aerial surveys, and concluded that volume scattering dominated the radar returned power in East Antarctica. Figure \ref{fig:MD} shows their predictions (yellow line). Overlaid with the Moore \& Davis prediction is the time evolution obtained in our own `fast' simulation (green points, described below). Moore \& Davis estimated a volume scattering contribution persisting for hundreds of ns beyond the initial surface reflection.
%MooreDavis.csv gnuplot

Yi and Bentley\cite{yi1994analysis} performed a similar analysis and reached a similar conclusion -- namely, that volume scattering typically accounts for approximately half of the returned signal power measured in the aerial surveys. In principle, there is a contribution to the total sub-surface reflected power from internal layers within the firn, however, typical measured layer reflectivities are of order -50$\to$ -70 dB\cite{besson2023polarization}, and therefore between 30$\to$50 dB weaker than surface echoes. Overall, these two calculations imply VS amplitudes significantly larger than what would be expected by a simple extrapolation of the Rayleigh formula. %However, owing to limited resources, the radar signals were never calibrated against a known reflector (e.g., aluminum sheet), which would have otherwise yielded a surface-only `standard', and allowed an absolute calculation of the sub-surface volume scattering cross-section.

%Recent work by Reference Ridley, and Partington,Ridley and Partington (1988) demonstrated that return wave forms from the ice sheets differed substantially from those predicted by the surface-scattering model in many cases. They proposed that volume-scattering from beneath the ice-sheet surface contributed substantially to the return power, and they developed a model based upon the numerical evaluation of an integral to describe the return wave forms. Based upon a qualitative analysis of averaged return wave forms, Reference Partington,, Zwally,, Ridley and RapleyPartington and others (1989) showed that the shape of the altimeter wave forms from Greenland corresponded roughly to surface-scattering in the low latitudes and volume-scattering in the higher latitudes. They observed only subtle variations in the shape of the averaged wave forms from the Antarctic plateau regions. Moreoever, Reference Davis, and Poznyak,Davis and Poznyak (in press) showed that 10 GHz radar signals can penetrate many meters below the ice-sheet surface of East Antarctica, and that the amount of penetration varied with surface elevation. Thus, sub-surface volume-scattering can contribute significant amounts of return energy at frequencies used by satellite altimeters.

Radio-frequency volume scattering in the icy moons of the 
gas giant planets of the Solar System has recently received renewed attention\cite{roberts2023exploring,romero2024feasibility}. The ice sheets of Europa, Ganymede, Enceladus, Miranda and Ariel e.g., are interesting not only from a hydrological perspective, but also as possible harbingers of viable life in the liquid oceans under their surface ice sheets. Those ice sheets have also been identified as candidates for measurements of radio emissions from neutrino-induced cascades\cite{gusev2010ice}. At the $\sim$100 K temperatures typical of those ice sheets, the RF attenuation lengths should exceed 10 km for pure ice, presenting the possibility of measurable echoes from the sub-surface ice/water boundary provided signal attenuation through the ice sheet itself is not prohibitively large. Calculating the thickness of the ice sheets from radar sounding therefore requires estimates of signal losses via VS. A recent calculation of the signal penetration associated with EM emissions from natural auroral activity, expected for the Uranian moons\cite{romero2024feasibility} found insignificant volume scattering losses, albeit at frequencies (f$\sim$100--900 kHz) much smaller than radio.%, for which the long-wavelength VS suppression would be expected to reduce such contributions to nearly negligible levels.

\subsection{~{\tt nuradiomc}-based Calculation and Modeling}
We have used the ${\tt nuradiomc}$\cite{glaser2020nuradiomc} simulation package to evaluate the expected impact of VS on neutrino waveforms. This Monte Carlo code is currently the most widely used software suite for following a signal ray from an interacting neutrino to an in-ice antenna. For our application, however, for which VS must be obtained by propagation of not only one signal ray, but a ray bundle over a volume growing as the cube of distance from the source, 
the superb fidelity offered by ${\tt nuradiomc}$ requires considerable CPU resources and limits the parameter space explored.

We vary the molecular radar cross-section in our simulation, resulting in varying amounts of simulated volume scattering. To reduce computational expense, we only track the first scatter and neglect secondary, tertiary, etc. scattering. Additionally, although both a direct (D) and refracted/reflected (R) ray typically connect a given source point to a given receiver, we limit the discussion below to the least-time (D) path, realizing that the volume scattering results presented below also apply to any R path, as well.

Additional parameters that must also be specified in the simulation include: i) polarization characteristics of VS (for which we assume that the atomic dipole re-radiates with a standard dipolar beam pattern relative to a given transmitter and retains the incident polarization), ii) coherence (either incoherent [IVS] or coherent [CVS]) of the re-scattered signal, and iii) the angular deviation $\delta\theta_C=\theta_C-\theta_{view}$ between the known in-ice Cherenkov angle of the emitted Askaryan radiation and the `viewing' angle relative to the axis of the simulated neutrino-induced shower, which is taken from the {\tt nuradiomc} event generator. The highest signal-to-noise ratio obtains when the in-ice radio receiver is positioned directly at the Cherenkov angle (viewing angle=$\theta_C$ and deviation off-angle $\delta\theta_C$=0), therefore, as $\delta\theta_C$ increases, the relative isotropic volume scattering contribution should increase. The admixture of signal-like CVS relative to noise-like IVS depends, in part, on the source$\to$receiver geometry.

For signal of initial amplitude $A_0$ in an infinite, uniform three--dimensional medium and separation distance between transmitter Tx and scatterer S equal to $r_{Tx,S}$, the amplitude of the signal $A_S$ arriving at time $t_S=r_{Tx,S}/c_0$ at the scatterer is proportional to $A_0/r_{Tx,S}$. In our {\tt nuradiomc} simulation, the Askaryan radiation signal, collimated into a Cherenkov cone of transverse width ${\cal O}$(1 degree) propagates to a distant radio receiver, separated by distances ranging from 50$\to$200 m. The volume surrounding the receiver is discretized into a three-dimensional cubical grid having cell volume 3000 ${\rm cm}^3$ (limited by CPU execution speed). Each volume element scatters the incident radiation losslessly, retaining the polarization of the incident signal; assuming dipole-like molecular scattering, the spatial distribution of scattered signal follows the standard $dN/d\theta\sim$sin$\theta$ dipole beam pattern. For both CVS and also IVS, we sum all the individual contributions, in appropriate time bins, to obtain the amplitude and time dependence of the scattered signal at the receiver. For each event, we perform a checksum and verify that the sum of the total scattered energy added to the unscattered energy matches the total calculated energy emission, in the absence of scattering.

Both CVS and IVS give an expectation value of receiver voltage $\langle V\rangle=0$, while the power delivered to a receiver ($P_{Rx}\sim\langle V^2 \rangle$) is positive-definite; IVS differs from thermal noise in that the source is considered to be a point in our simulation rather than a three-dimensional environmental background source, which results in a characteristic time dependence of the IVS power envelope. %\textcolor{red}{Ray bending due to the variable refractive index (for which we use a model n(z)=1.78-0.438exp(0.016z), with z the ice depth in meters\cite{Allison:2019rgg}) is already incorporated into the simulation.} 
Our simulation assumes that both Tx/Rx are sufficiently deep that the surface ice-air boundary can be neglected. To reduce computing time, the simulation also assumes that any possible inhomogeneities in the ice are much smaller than the characteristic wavelength scale. %For constant density, the total incoherent signal amplitude measured at the receiver results from competition between the 1/r dependence of the signal, and the $r^2$ growth of a spherical shell contributing signal to the receiver, such that the number of scatterers N in a given shell of radius r $N\propto r^2$. 

\begin{figure}
\centerline{\includegraphics[width=\textwidth]{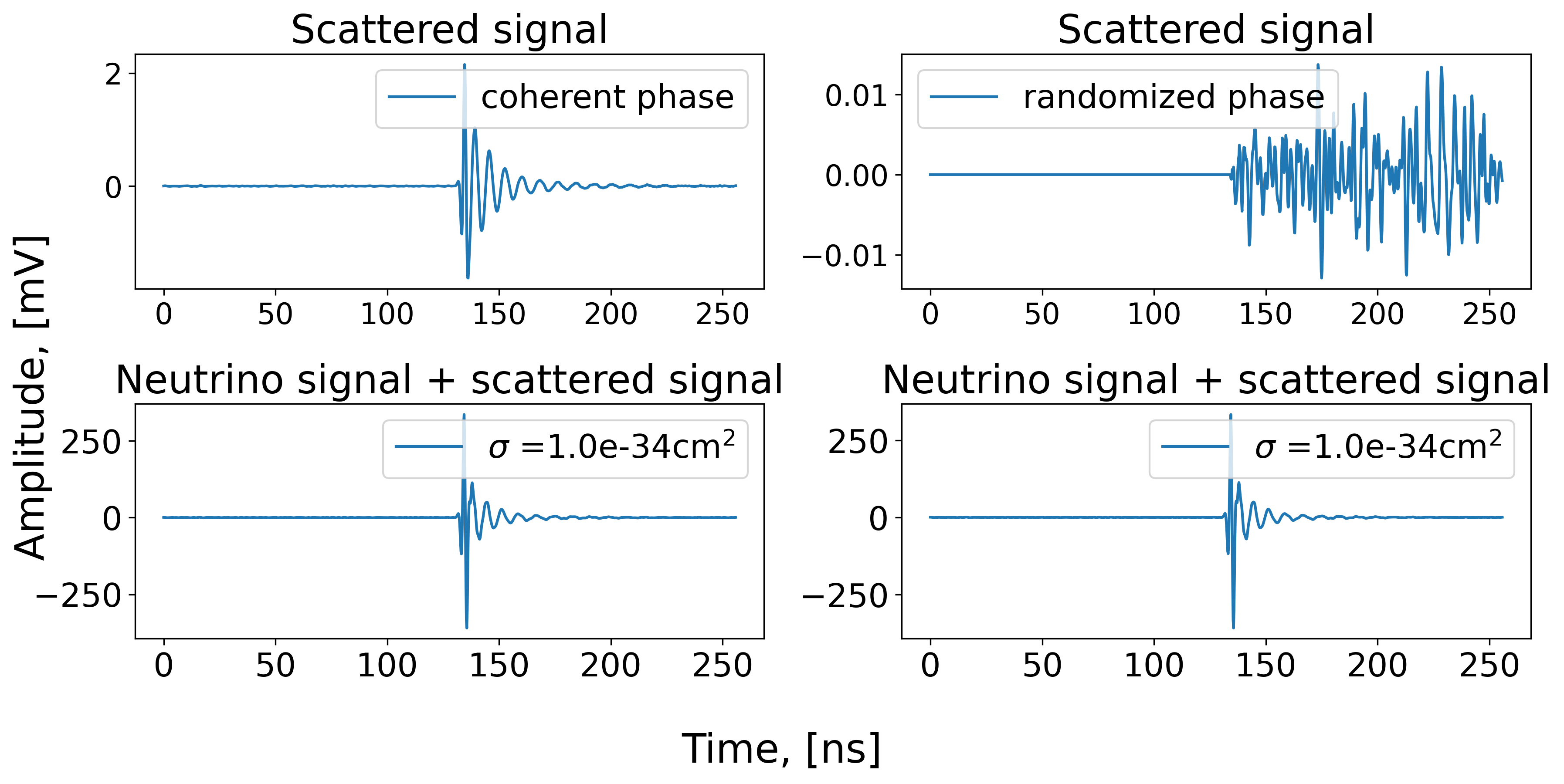}}
\caption{{\tt nuradiomc} simulations showing original on-cone Askaryan signal compared to either CVS (upper left, and also superimposed on the Askaryan signal in the lower left) or IVS voltage profiles (upper right, and also superimposed on the Askaryan signal in the lower right) expected for upper limit VS scattering cross-section ($1\times 10^{-34}~\text{cm}^2$) and also corresponding composite (Askaryan signal + VS contributions) waveforms. In both cases, for this value of separation distance (r=50 m), modification to original signals is immeasurably small as evident from the lower two panels, so the received waveforms are visually indistinguishable. As expected, coherent summation of VS Huygens wavelets leads to a larger amplitude than for the incoherent case, as the electric field dependence with distance should vary as 1/r in the former case and $1/r^2$ for the latter.}
\label{fig:SignalModeling}
\end{figure}

We obtain from the simulation both the relative shape, as well as the absolute normalization of the voltage profile for either coherent (CVS) or incoherent (IVS) volume scattering as a function of time relative to the initial signal onset resulting from the in-ice neutrino interaction, for a given scattering cross-section.  
Figure \ref{fig:SignalModeling} shows the results of our simulation for both CVS and IVS. CVS produces, for this 50 m separation distance, a roughly 1\% increase in the measured amplitude with duration only slightly longer than the primary signal. IVS produces a $<$0.1\% increase in the featureless noise following the signal onset. 

Once again referencing the experimentally measured polar ice attenuation length of \mbox{800 m}, we first use our simulation to obtain a refined upper limit of the VS cross-section, to be compared with the dimensional analysis presented previously. Here, we vary the scattering cross-section until the signal loss, beyond standard 1/r amplitude reduction, accounts for the 800 m field attenuation length typical of in-ice radio experiments. We obtain an upper limit on the scattering cross-section of $1\times 10^{-34}~{\rm cm^2}$ from this exercise, and take this to be the current level at which the scattering cross-section is constrained by extant data. 
Scanning over cross-sections from $10^{-35}-10^{-30}$ $\text{cm}^2$ in our simulation yields the result that the CVS amplitude scales roughly quadratically with cross-section. % (Figure \ref{fig:ACVS_v_XSct}). 
%At sufficiently high cross-sections, propagating electromagnetic waves are scattered well before reaching the region of the receiver, thereby mitigating the remaining VS power.
%\begin{figure}    \centering    \includegraphics[width=0.75\linewidth]{apix/ACVS_v_XSct.png}    \caption{CVS amplitude, as a function of input Volume Scattering cross-section, determined from ${\tt nuradionc}$ simulations.}    \label{fig:ACVS_v_XSct}\end{figure}

\begin{figure}
\centerline{\includegraphics[width=\textwidth]{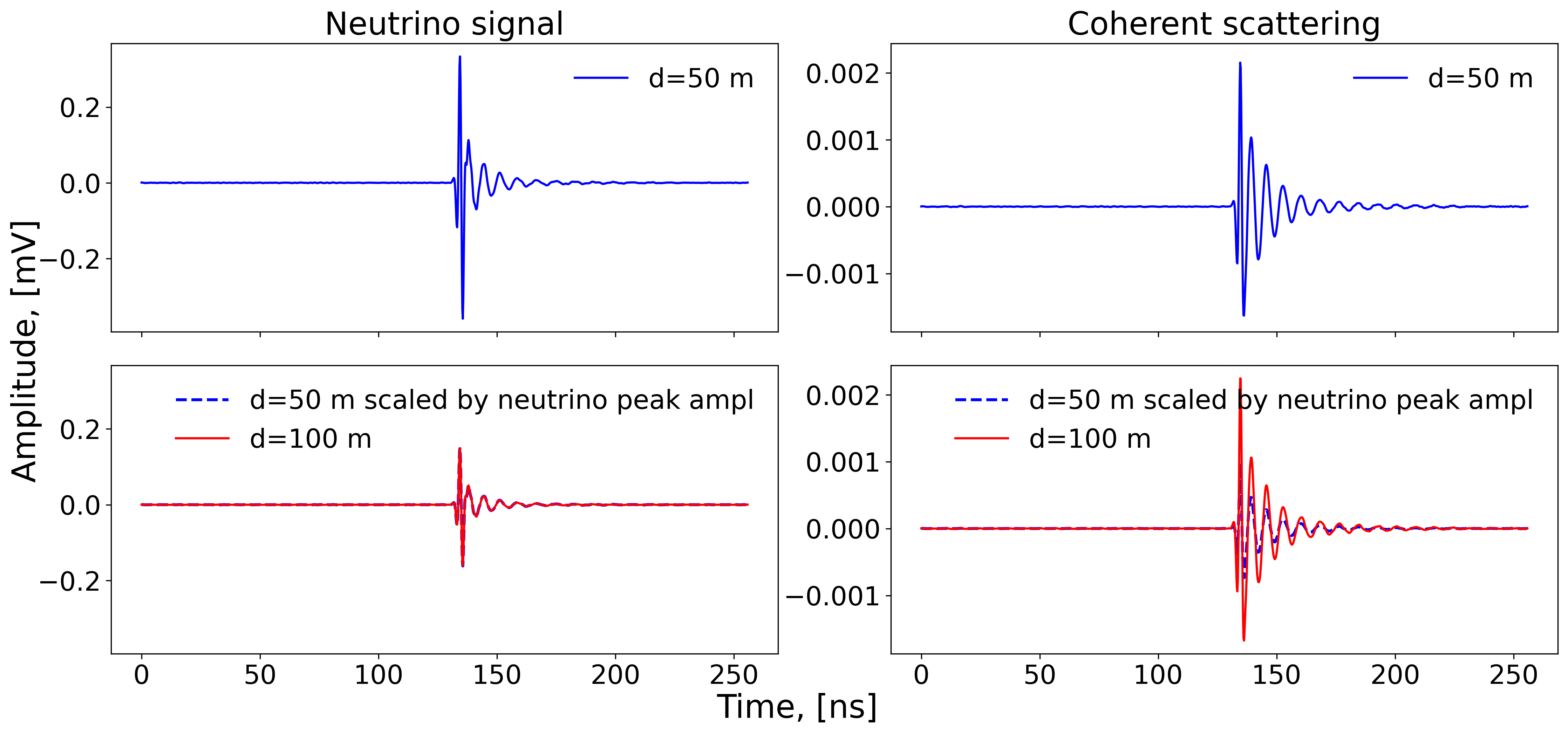}}
\caption{Comparison of on-Cherenkov-cone signal amplitude at r=50 m (top, left) and r=100 m (bottom, left), assuming maximal scattering cross-section ($\sigma=1\times 10^{-34}~\text{cm}^2$) allowed by current ice attenuation length measurements. Right panels show corresponding simulated CVS voltage profiles at specified (source, receiver) separations. Note that the CVS contribution increases more rapidly than 1/r, owing to increasing flux collimation as the distance between Tx and Rx increases.}
\label{fig:CVSvsR}
\end{figure}

\begin{figure}\centerline{\includegraphics[width=0.83\textwidth]{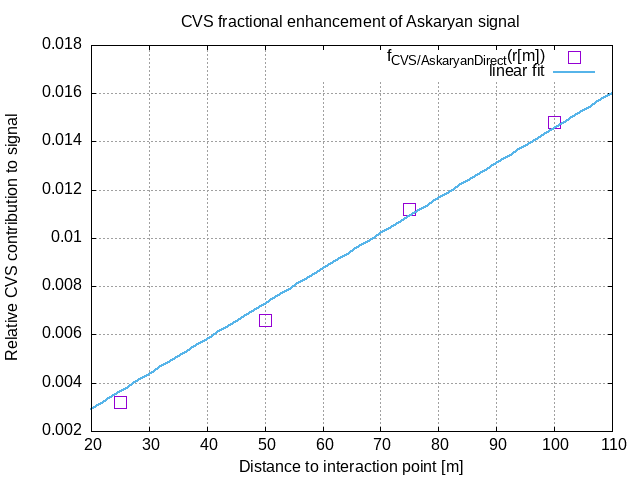}}\caption{Ratio of CVS amplitude to on-cone Cherenkov signal strength, using same parameters as in previous plot ($\sigma=1\times 10^{-34}~\text{cm}^2$), as a function of transmitter-receiver separation distance. Overlaid (blue line) is an approximate linear fit.}\label{fig:CVSvsRtrend}\end{figure}

For a given cross-section, we can now evaluate the impact VS might have on observed neutrino signals.
Since the neutrino signal arrives at the receiver from a variety of distances and angles, it is important to assess how any putative VS may similarly vary with range and event geometry.
Figure \ref{fig:CVSvsR} shows the relative CVS contribution, on the Cherenkov cone, for r(Tx,Rx)=50 m and also r(Tx,Rx)=100 m. 
As the separation distance increases, the CVS flux bundle which contributes power at the location of the receiver close to the signal onset time geometrically constricts, resulting in an increasing density of flux lines at the receiver. The CVS amplitude correspondingly falls off less rapidly than the Askaryan signal with increasing distance. 
%Figure  shows
With our simulations, we have calculated the fractional enhancement f of the peak neutrino signal amplitude due to CVS, assuming the maximal scattering cross-section ($\sigma=1\times 10^{-34}~\text{cm}^2$) derived above. We observe an approximately linear increase in the relative CVS contribution, with f(r)$\sim 1.6\times 10^{-4}$r[m] (\autoref{fig:CVSvsRtrend}). To verify this trend at larger values of separation distance, a `fast' simulation, which is limited to two-dimensions and eschews all frequency-dependent signal/receiver details as well as ray tracing, was written; that faster simulation qualitatively predicts a similar increase in background amplitude with distance, on-cone, out to multi-kilometer distance scales.  %such that at a neutrino range of 1.5 km, the extrapolated signal fraction approaches 10\%, under these extreme assumptions.

\begin{figure}
\includegraphics[width=0.5\textwidth]{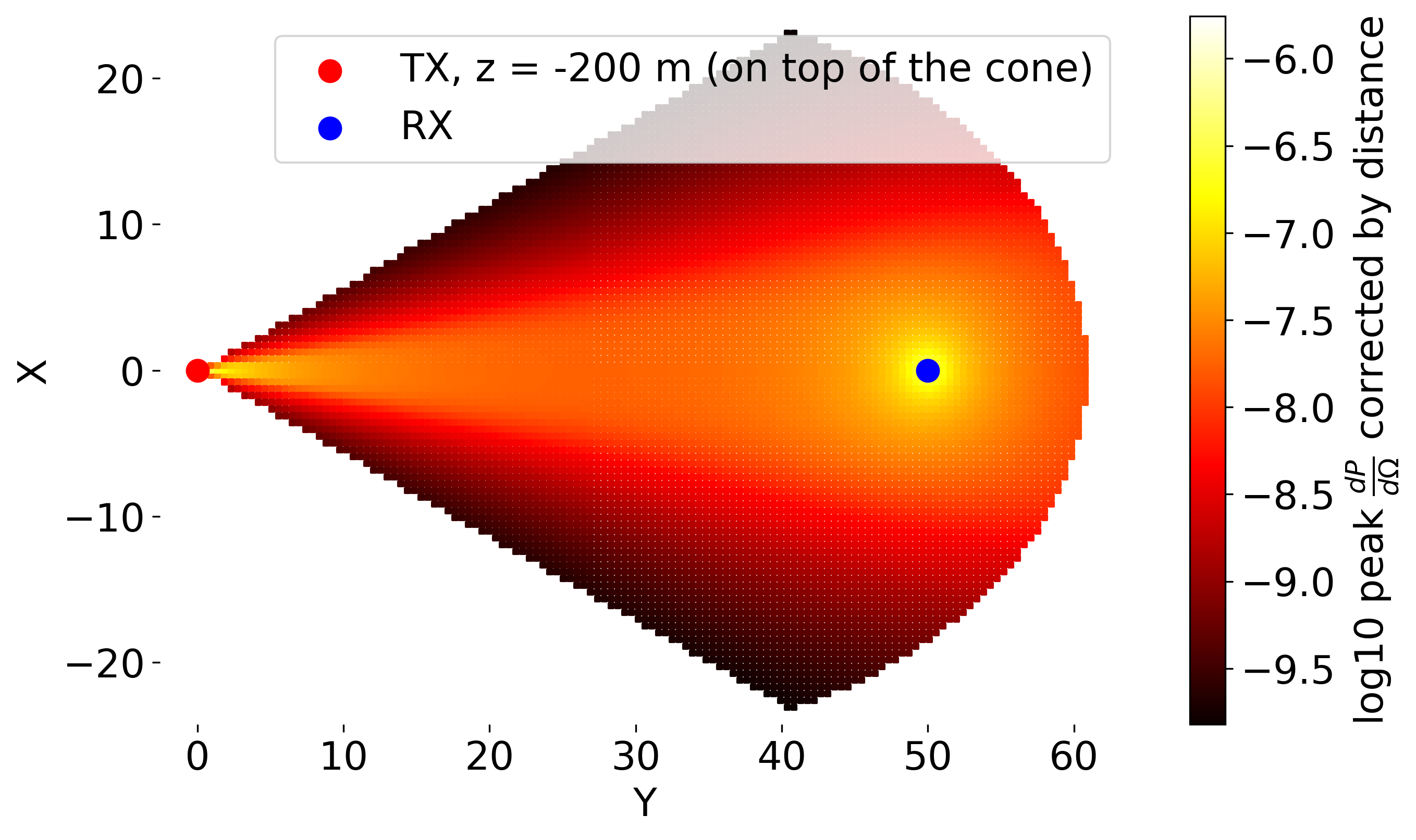}
\includegraphics[width=0.5\textwidth]{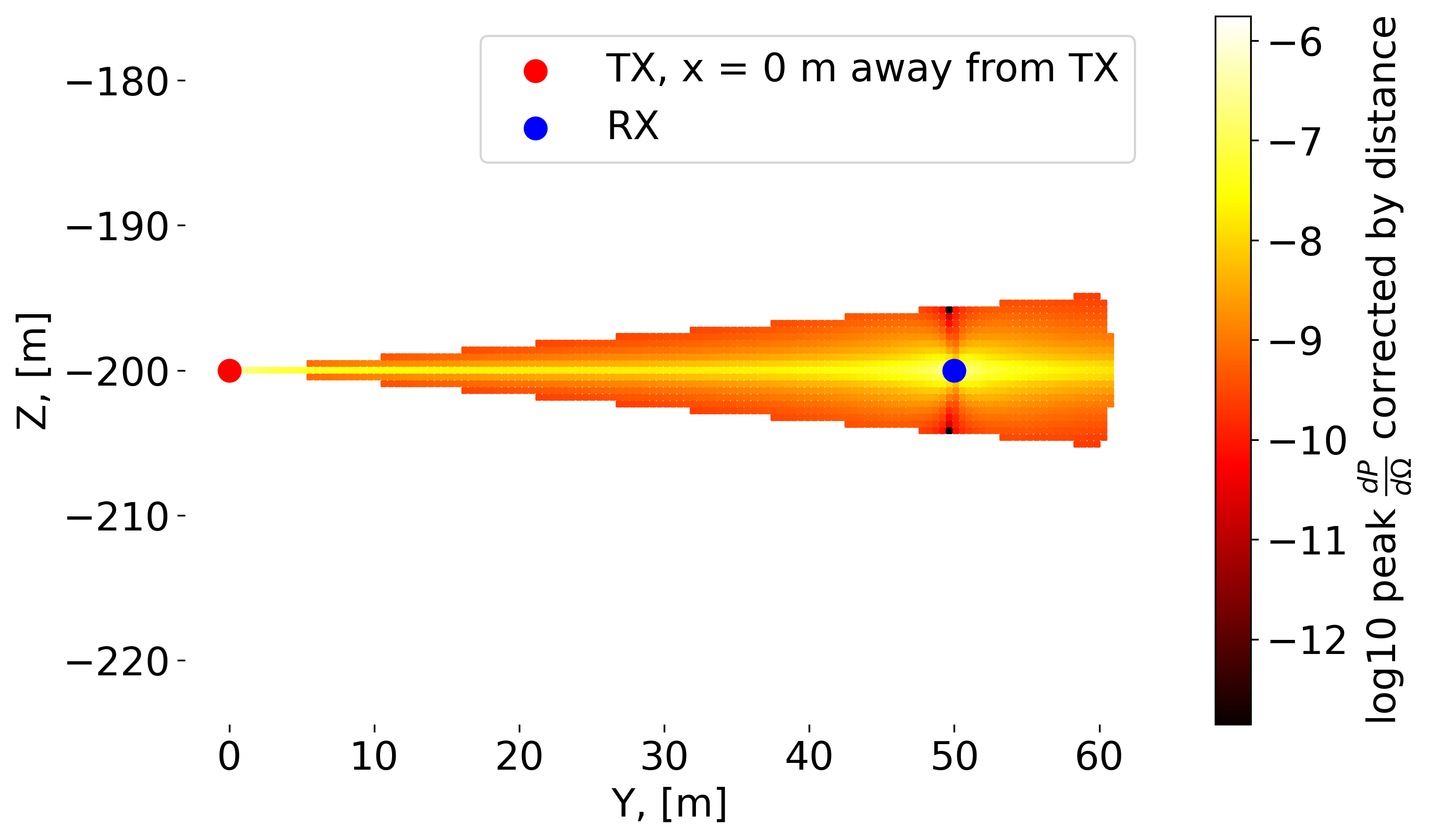} \\
\centerline{\includegraphics[width=0.8\textwidth]{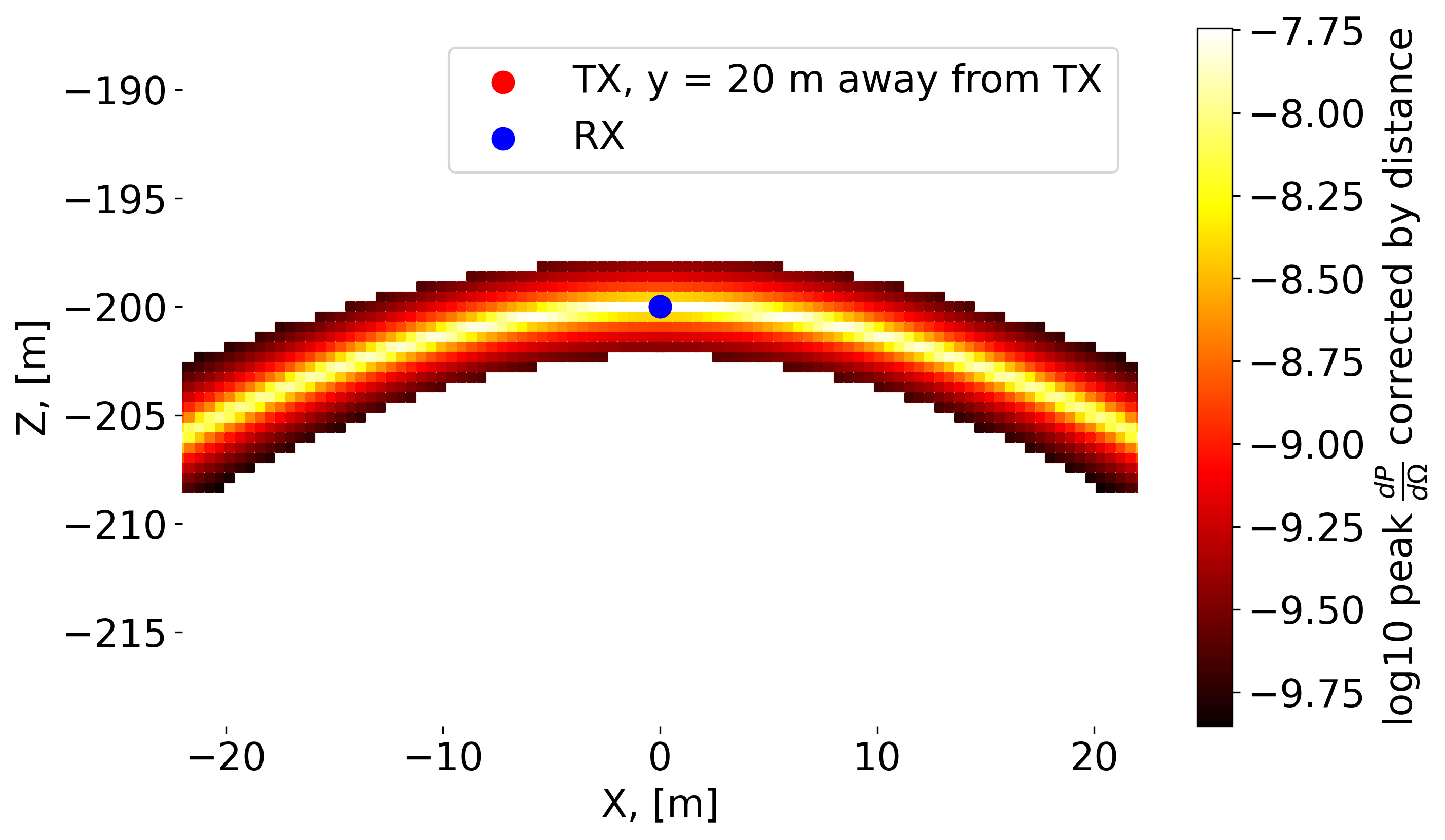}}
\caption{Three planar projections of simulated Cherenkov cone. Shower cone is centered on y-axis; neutrino momentum vector points in +y direction. Shown are xy- (top left), yz- (top right), and xz-planar projections. For these simulations, the receiver (blue circle) has been located at a position corresponding to maximal illumination of the Askaryan signal. As the receiver location is moved off the Cherenkov cone, however, the relative ratio of the volume-scattered amplitude relative to the direct Askaryan signal begins to increase.}
\label{fig:CCgeo}
\end{figure}

Since the Askaryan signal is confined to a cone of transverse width $\sim$1-2 degrees, the fractional VS contribution (relative to direct Askaryan) is expected to grow as the receive (viewing) angle increasingly deviates from the Cherenkov angle. Figure \ref{fig:CCgeo} illustrates the Cherenkov signal geometry in three planar cross-sections, for the case where the receiver (black dot) is maximally illuminated. Since the signal is so tightly collimated in angle (the 2 degree FWHM is illustrated in the bottom panel of \autoref{fig:CCgeo}), as the receiver is increasingly displaced off-cone, the VS contributions from volume elements remaining on-cone will proportionately increase as the Askaryan neutrino signal rapidly weakens. 

\begin{figure}
    \centering
    \includegraphics[width=1\linewidth]{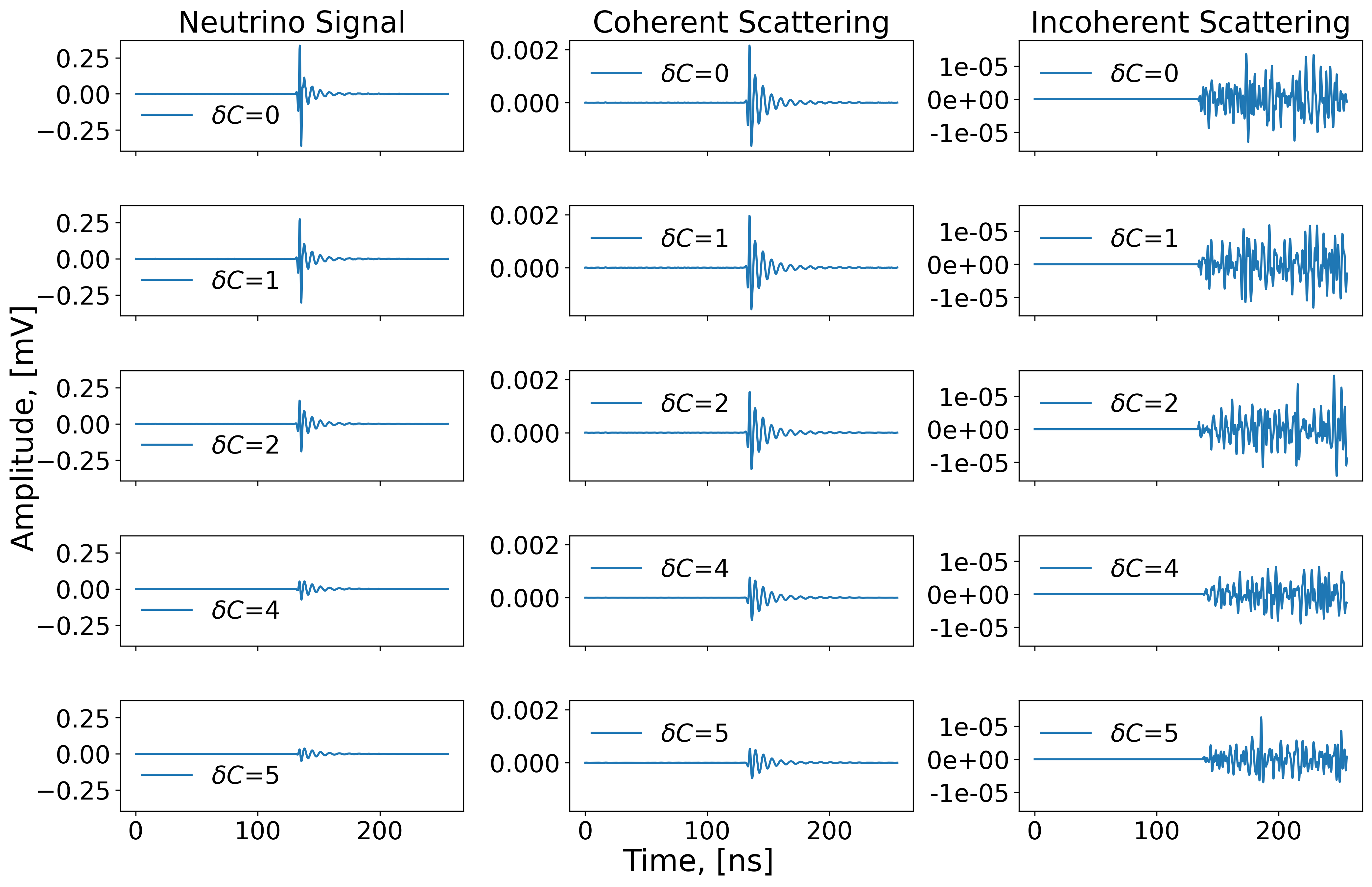}
    \caption{Simulated direct neutrino-induced Askaryan signal (left column), CVS signals (center column) and IVS signals (right column) at 100 m separation between Tx and Rx, as a function of the viewing angle, defined as the angular deviation between a receiver and the Cherenkov angle $\delta$C (in degrees). As $\delta$C increases, the signal-like CVS magnitude increases relative to the direct Askaryan signal, whereas IVS simply adds noise power to the background `under' the signal.}
    \label{fig:enter-label}
\end{figure}
\begin{figure}
\centerline{\includegraphics[width=0.8\textwidth]{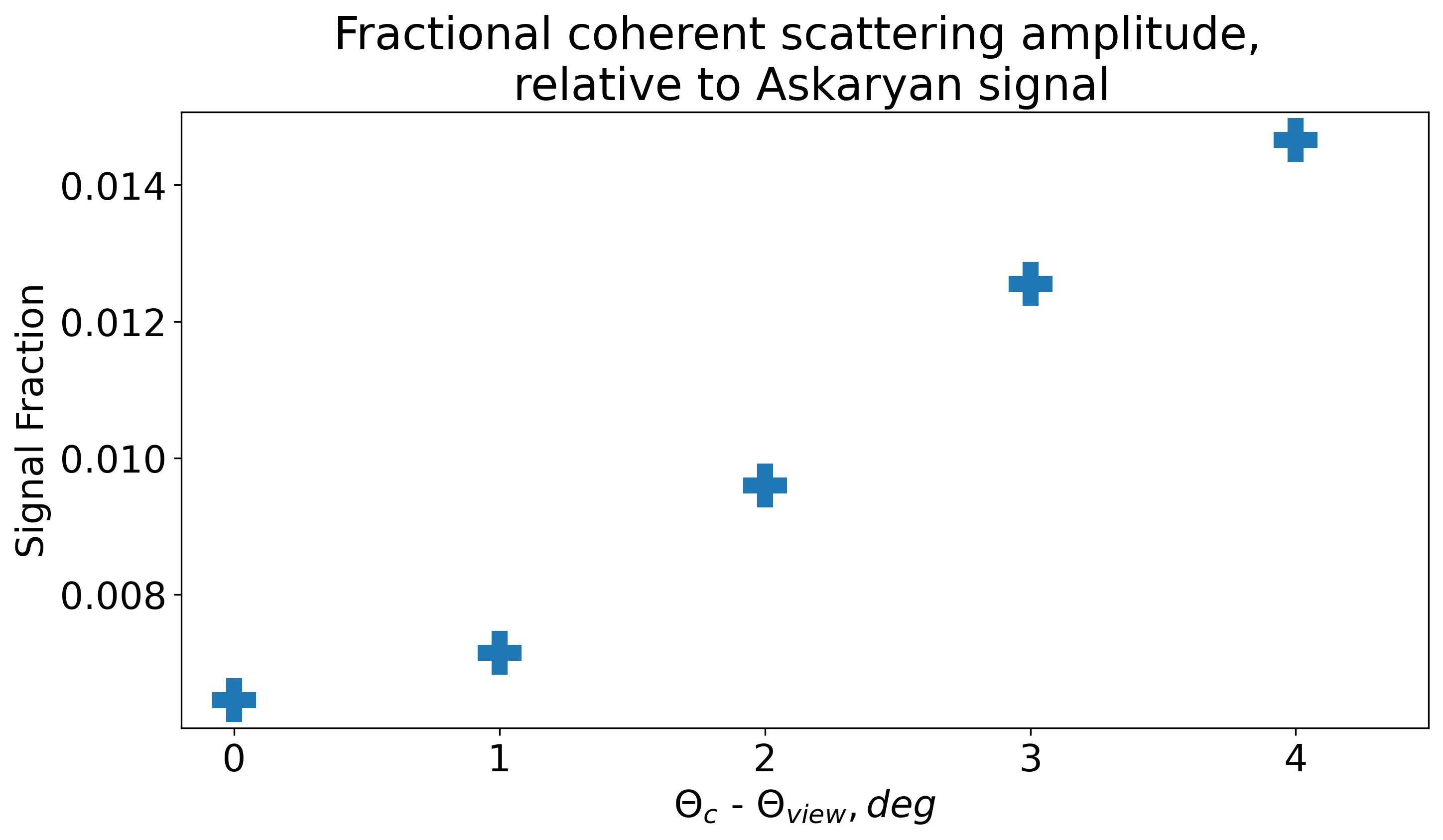}}
\caption{Ratio of CVS amplitude to off-cone Cherenkov signal strength, at separation distance of 50 meters, as function of viewing angle.}
\label{fig:CVSoffCone}
\end{figure}

\autoref{fig:enter-label} displays simulated waveforms as the viewing angle is increased from on-cone to a deviation of 5 degrees, illustrating the proportional increase in the relative VS fraction. Based on those waveforms,
Figure \ref{fig:CVSoffCone} shows the relative CVS amplitude, as the received Askaryan signal deviates from the Cherenkov angle. The CVS contribution decreases only very slowly with angle, while the Askaryan signal falls as an exponential with characteristic scale of approximately 1 degree. 

\section{~Incoherent Volume Scattering Experimental Data Analysis}
\begin{figure}
\centering
\includegraphics[width=0.8\textwidth]{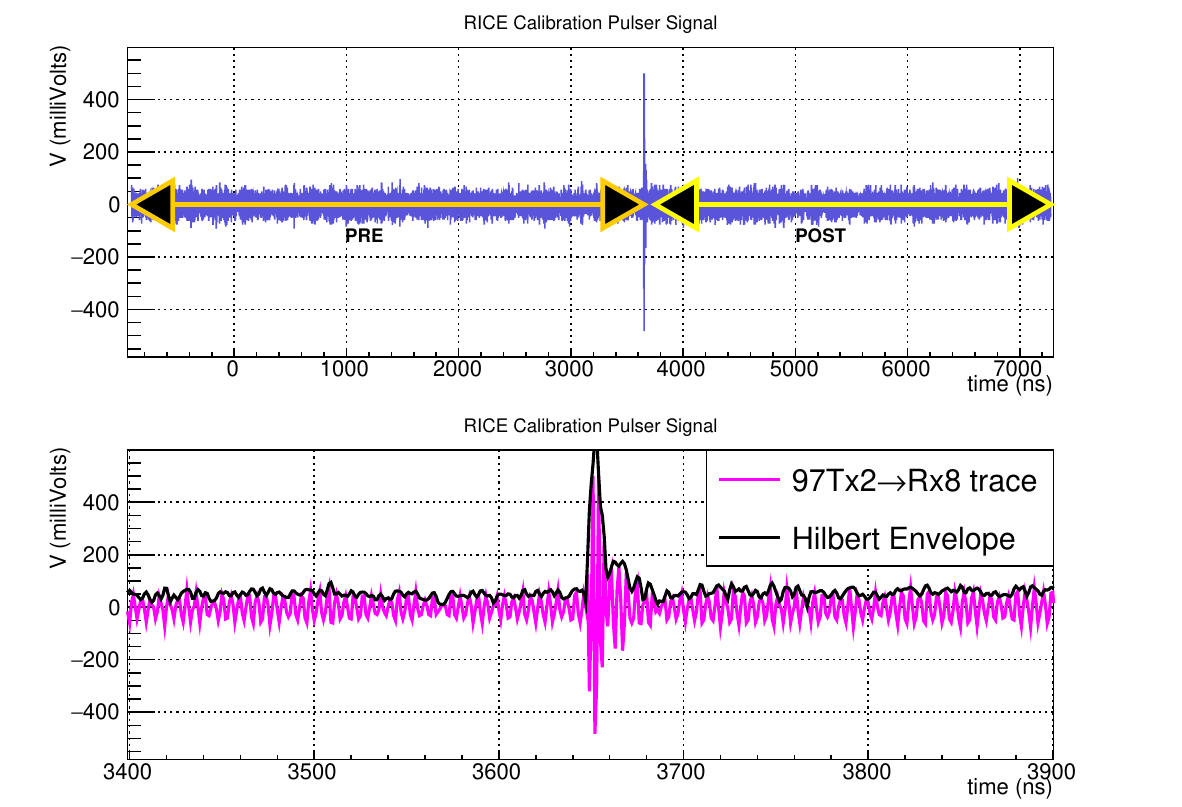}
\caption{Top: Typical RICE calibration pulser signal; `PRE' and `POST' time interval definitions are as indicated in the Figure, and are used to calculate possible excess IVS contribution in POST time interval. Bottom: Zoom of time interval around peak with Hilbert Envelope overlaid; we observe that signal ringing persists for up to 50 ns owing to dispersive effects in transmitter and receiver signal chain.}
\label{fig:RICECP}
\end{figure}

\begin{figure}
\centerline{\includegraphics[width=\textwidth]{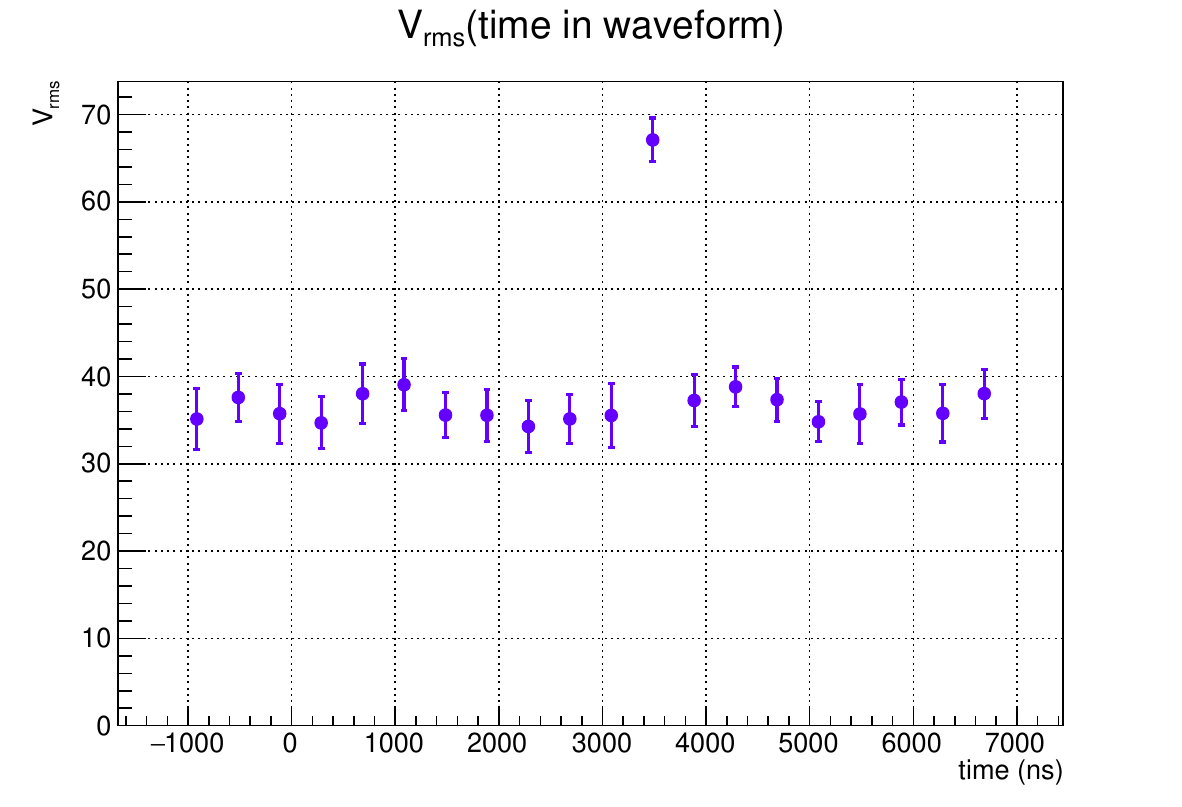}}
\caption{Voltage rms, calculated from rebinning previous waveform in twenty 400-ns bins, as a function of depth into waveform capture. We observe that the rms for the POST-signal region is approximately equal to the PRE-signal region, contrary to the expectations if there were a large IVS contribution.}
\label{fig:RICEcrms}
\end{figure}

We use experimental in-ice calibration pulser data from the RICE experiment\cite{KravchenkoFrichterSeckel2003,Kravchenko:2011im} to study possible volumetric scattering effects. 
Since the CVS waveform is so similar to the direct signal, experimental separation of the two is correspondingly difficult, and we therefore focus on the IVS component only. Selecting events with SNR values typical of neutrinos,
we quantify any possible incoherent volume scattering component produced by the local calibration pulser by subtracting
the PRE-signal portion of the waveform from the POST-signal portion of the waveform, and calculate the magnitude of any POST excess. 
 From our simulations, we expect that IVS would result in superposition of the waveforms shown in the upper panels of \autoref{fig:SignalModeling} on the expected brief, $\sim$20-30 ns impulsive signal peak, producing an excess of average noise power after (``POST") the calibration pulser signal is received, relative to before (``PRE") the signal arrival. Transmitter signals were broadcast to the receiver antennas (approximately 100 meters distant) that comprised the RICE radio receiver array. A typical received pulse (Figure \ref{fig:RICECP}) illustrates the intrinsic antenna response, consisting of a `ring-down' of duration tens of ns, evident in the lower panel of \autoref{fig:RICECP} (including any possible coherent volume scattering component within that time scale), superimposed on the incoherent background from thermal noise (constant in time) and any possible additional IVS. Rebinning the waveform to allow a coarse tracking of the rms voltage(time), we observe that the rms is approximately equal in the POST and PRE time intervals (\autoref{fig:RICEcrms}), disfavoring any apparent IVS component.

%We assume that the emitted pulses are identical, such that observed received pulses differ primarily in the fluctuations of the background to the broadcast signal. 

%Using a model for the time evolution of the CVS or IVS signal power allows us to fit the observed time profile of the captured waveforms and extract the magnitude of any possible volume scattering component.

%\item Pulsed, and also Continuous wave (CW) `tone' data from the Radio Ice Cherenkov Experiment, originally collected in 2004 to quantify the radio-frequency attenuation length can also be investigated for experimental evidence of volume scattering. A CW signal generator was gated to yield signal durations ranging from 50 nanoseconds to several microseconds, at frequencies ranging from 211--491 MHz, which also allows comparison of a possible volumetric scattering component with the expected $1/\lambda^4$ Rayleigh scattering.\end{itemize}

\begin{figure}
\centerline{\includegraphics[width=\textwidth]{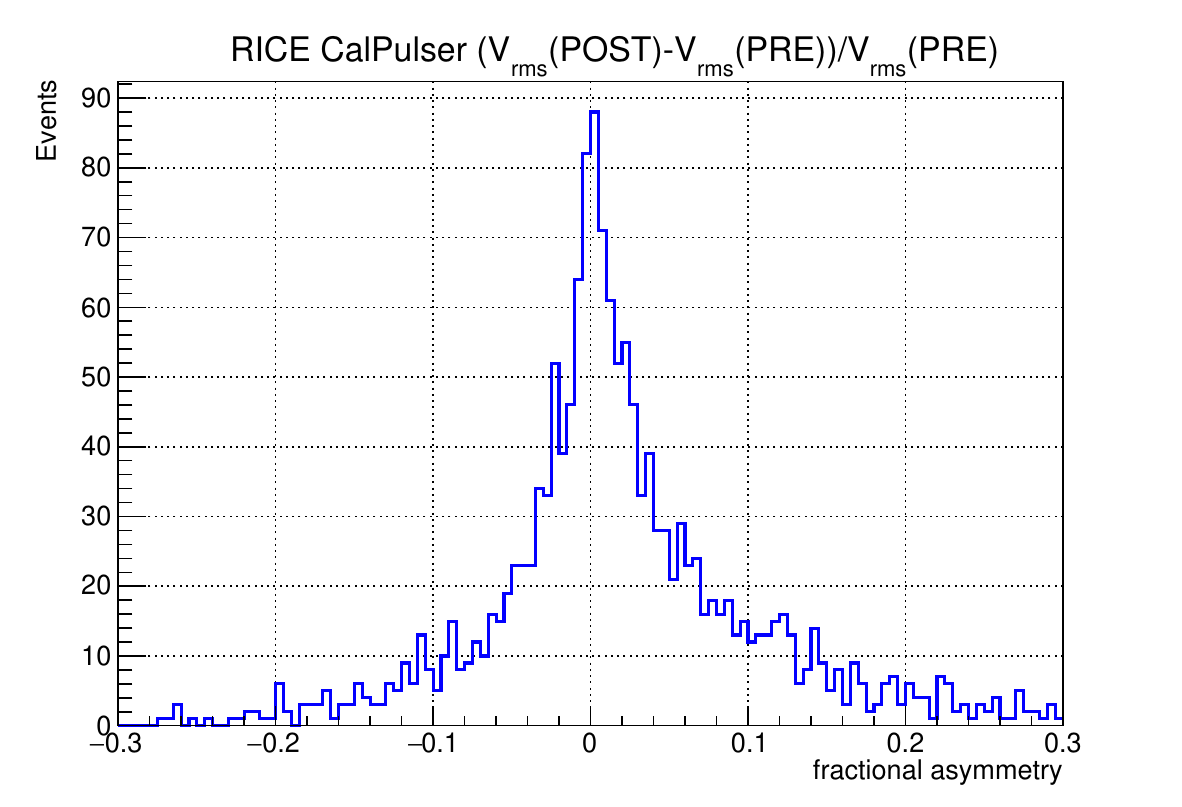}}
\caption{Distribution of excess incoherent contribution to Hilbert Envelope (POST-signal), relative to PRE-signal thermal noise for local calibration pulser event captures based on approximately 200 events similar to \autoref{fig:RICECP}. The distribution closely follows a Gaussian; the slight offset in the mean from zero (0.18\%) may be an artifact of antenna `ringing'.}
\label{fig:RIVS2TN}
\end{figure}

\autoref{fig:RIVS2TN} shows the distribution that obtains by analyzing approximately 200 such events, and presents the (POST-PRE)/PRE fraction as a measure of the total IVS component, relative to thermal noise. We note a slight positive offset, relative to zero, which may also be due to inclusion of the `tail' of the antenna ring in our POST summed amplitude.
This distribution corresponds to a negligible added signal power corresponding to 0.18\% of the thermal noise power, at an experimental source/Rx separation distance of 140 meters.

%We note that the signals from vertically-polarized dipole calibration pulsers illuminate all elevation angles, following the standard dipolar $A(\theta)\sim\cos\theta$ beam pattern. Askaryan signals, by contrast, are collimated into a \mathcal{O}(1-2 degree) Cherenkov cone. 

An additional experimental handle could be afforded by the POST/PRE signal shape dependence on the local density of the environment -- volume scattering for shallow antennas should be reduced relative to deep antennas, owing to the smaller local density. 
\begin{figure}
\centering
\includegraphics[width=0.8\textwidth]{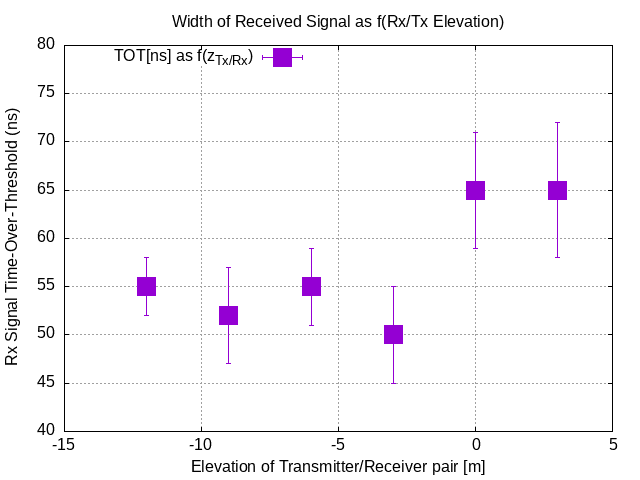}
\caption{Width (`Time-Over-Threshold', or `TOT', in ns) of Hilbert envelope as function of depth of transmitter/receiver pair, for waveforms recorded as transmitter/receiver dipole pair were co-lowered into South Polar ice. In general, the shape of the signals track each other, albeit with some broadening for the in-ice case, with no evident additional contribution from VS. Surface reflection has not been subtracted from +3 m elevation data point.}
\label{fig:B2XB4}
\end{figure}
Figure \ref{fig:B2XB4}, for example, shows the width of the Hilbert envelope of the recorded waveforms, as a RICE dipole transmitter/receiver pair are co-lowered into neighboring boreholes, laterally separated by 30 m. At an elevation of +3 m, we assume an ambient refractive index n=1.0; at z=-12 m, the refractive index is approximately 1.45, corresponding to ice density approximately one-half of the maximum asymptotic value. Overall, we observe no extension of the signal width, as would be expected from CVS contributions; in fact, the signal appears to only narrow with depth. Interestingly, the signal arrival time is observed to be approximately equal for an elevation of +3 m vs. zTx=0 m (vertically-oriented Rx/Tx dipoles both halfway into the ice), indicating at least one exclusively in-air path for that case. 

\section{~Summary, Conclusions and Future Work}
Although often referenced in radioglaciological literature, the polar ice volume scattering cross-section has, thus far, lacked quantification. Experimental VS constraints are particularly important for efforts seeking measurement of radio emissions from in-ice neutrino interactions. Herein, we have derived an upper limit on the VS cross-section using bistatic radar echo data, modeled the expected CVS and IVS signatures using the {\tt nuradiomc} simulation package and also derived limits on the incoherent volume scattering contribution to the ambient background level, relative to measured thermal noise, using data taken with the RICE experiment. Given our estimated cross-section limit, coherent volume scattering may result in an increase of up to 10\% in the magnitude of the measured neutrino-induced signal (depending on event geometry) amplitude, with a small distortion in the voltage vs. time profile. From data, we find that the maximum allowed contribution of IVS is of order 1\% relative to the mean thermal noise background.
These results are relevant to detection of radio-frequency signals generated by collisions of neutrinos with ice molecules, such as with the current ARA, ARIANNA, PUEO, RET, and RNO-G experiments, and the planned radio component of the IceCube-Gen2 experiment.
 Future, more stringent bounds on VS may be possible with additional calibration pulser data taken over a wider range of transmitter-receiver ranges and incidence angles, and/or an improved understanding of how to properly evolve precise antenna response measurements made in-air to dense media such as ice. Particularly useful would be detailed comparisons of transmitter signals broadcast in-air to transmitter signals broadcast in-medium, from which both CVS and IVS contributions may be extracted, as a function of the local ice density. 

\section{~Acknowledgements}
This work was made possible, in part, by the National Science Foundation's generous IceCube EPSCoR Initiative  grant \#2019597.

\bibliographystyle{JHEP}
%% \bibliography{biblio.bib}

\bibliography{igsrefs}
\end{document}